\documentclass[%
 reprint,
amsmath,amssymb,
aps, 
]{revtex4-2}

\usepackage{graphicx}
\usepackage{dcolumn}
\usepackage{bm}
\usepackage{comment}

\setcitestyle{super}

\begin{document}

\preprint{PRN/123-QED}
\title{
Sub Band Gap Operation Limits for Perovskite Light Emitting Diodes
}

\author{Pradeep R. Nair}
 \email{prnair@ee.iitb.ac.in}
\affiliation{ Department of Electrical Engineering,
 Indian Institute of Technology Bombay, Mumbai, Maharashtra, India 400076\\
}%




\date{\today}

\begin{abstract}
Ultra low voltage operation for Perovskite light emitting diodes (PeLEDs) has been demonstrated in recent years as high radiance with minimal power consumption is a desired feature. However, the light output at such conditions from PeLEDs is typically very low, and the maximum in external quantum efficiency (EQE) and energy conversion efficiency (ECE) are achieved at large biases with significant power consumption. Here, we explore the possibility of achieving maximums in EQE and ECE at sub band gap voltages for PeLEDs. Our analysis consistently interprets otherwise scattered experimental data from literature, identifies the limits for low voltage operation, and elucidates optimization routes for sub band gap high radiance operation of PeLEDs. 
\end{abstract}

\keywords{power consumption, Joule heating, quantum efficiency, }
\maketitle

\section{\label{sec:intro}Introduction}

The last decade has witnessed impressive performance gains in Perovskite light emitting diodes (PeLEDs)  with excellent EQE and improved stability\cite{li2024high,guo2022ultrastable,yuan2024bright, Li2024,song2025intragrain,baek2024grain,ding2024phase,ma2025recent,ke2025high,kim2022ultra}. For energy efficient applications, it is important to achieve high radiance at low power consumption. Accordingly, ultra low voltage operation of PeLEDs has attracted much recent research interest \cite{lian2022ultralow,zheng2024ultralow,wang2025efficient}. Interestingly, light emission has been observed from PeLEDs at sub $E_g$ voltages, where $E_g$ is the band gap of active material\cite{lian2022ultralow}. 
While detection of light emission at ultra-low voltages is limited by the sensitivity of optical measurement systems, it is necessary to achieve significant emission at low voltages to minimize power consumption in light emitting diodes (LEDs). Despite the impressive performance improvements, PeLEDs are yet to achieve maximum EQE or ECE at sub $E_g$ voltages. Further, the literature lacks explicit theoretical guidelines for performance optimization towards high radiance low voltage operation of PeLEDs. \\

A review of the recent literature on ultra low voltage operation of PeLEDs leads to the following open questions: (a) What are the ultimate limits for low voltage operation of PeLEDs? (b) Why does the turn ON voltage scale linearly with $E_g$ with a near constant offset of about $1\,\mathrm{V}$ for several LED technologies\cite{lian2022ultralow}? (c) Is it possible to achieve maximum EQE and ECE under sub $E_g$ applied biases? (d) The literature\cite{Li2024,jia2021excess,zhao2020thermal,zheng2024ultralow,wang2025efficient} indicates that the applied biases at which EQE maxima are achieved seem to be independent of $E_g$. Is this an expected trend? and (e) What are the optimization pathways to achieve high radiance low voltage operation for PeLEDs? In this manuscript, we address these open questions through well calibrated analytical models. Our model identifies the limits for sub $E_g$ operation of PeLEDs and accurately accounts for several recent observations which include the near $1\,\mathrm{V}$ offset between the $E_g$ and turn on voltage. Further, we identify the optimization routes to achieve maximum EQE and ECE at sub $E_g$ voltages. These insights are of broad interest toward high luminance, low power operation of PeLEDs (and for other LEDs as well).\\

Below, we provide a concise description of the current-voltage (JV) characteristics of PeLEDs which helps to identify the limits for ultra-low voltage operation of PeLEDs.
\section{Analytical model}
PeLED characteristics like the dark JV, EQE, ECE, and radiance could be predicted through a multi-physics approach. Under the assumption of spatially invariant carrier densities $n=p$, PeLED characteristics can be described by the following equations\cite{nair_jpcl2014,Hossain2024,nair2025multiphysics}:
\begin{align}
    J & = q \left(k_1n+k_2n^2+k_3n^3\right)W_P 
    \label{eq:Jrec}\\
    n&=n_ie^{qV_D/2k_BT}
    \label{eq:VD}\\
     V_{SC}&=K_{SC}J^\alpha 
\label{eq:Vsc}\\
     V_{app} & =  V_D+V_{SC} 
     \label{eq:Vapp}
\end{align}
Here, $J$ is the current and $V_{app}$ is the applied bias. The current is assumed to be recombination limited with monomolecular, bimolecular, and Auger components with $k_1$, $k_2$, and $k_3$ being the respective coefficients and $W_P$ is the active layer thickness (see eq. \ref{eq:Jrec}). The voltage drop in the active layer $V_{D}$ modulates the carrier density as per eq. \ref{eq:VD} where $n_i$ is the intrinsic carrier concentration, $k_B$ is the Boltzmann's constant and $T$ is the temperature. The applied voltage $V_{app}$ supports both the recombination and space charge limited transport, if any (described by eqs. \ref{eq:Vsc}-\ref{eq:Vapp}). Here, $k_{SC}$ and $\alpha$ characterize the space charge limited transport. This model can be appropriately modified to account for shunt and series resistances, if any. Note that the assumption $n=p$ is, in general, not valid for classical PIN diodes. However, the same is valid in the presence of mobile ions and hence appropriate for perovskite based optoelectronic devices like solar cells and diodes\cite{reenen_jpcl,saketh2021ion}.\\

The net photon flux from the LED is given as 
\begin{equation}
 N_{ph} = k_{2}n^2\times W_P\eta_{OC} 
 \label{eq:Nph}
\end{equation}
where  $\eta_{OC}$ is the outcoupling factor. The performance parameters are given as
\begin{align}
    \mathrm{EQE} & = \frac{N_{ph}}{J/q}
    \label{eq:eqe}\\
     \mathrm{ECE} & = \frac{qE_gN_{ph}}{JV_{app
     }} 
     \label{eq:ece}
\end{align}

\begin {figure} [h!]
  \centering
    \includegraphics[width=0.45\textwidth]{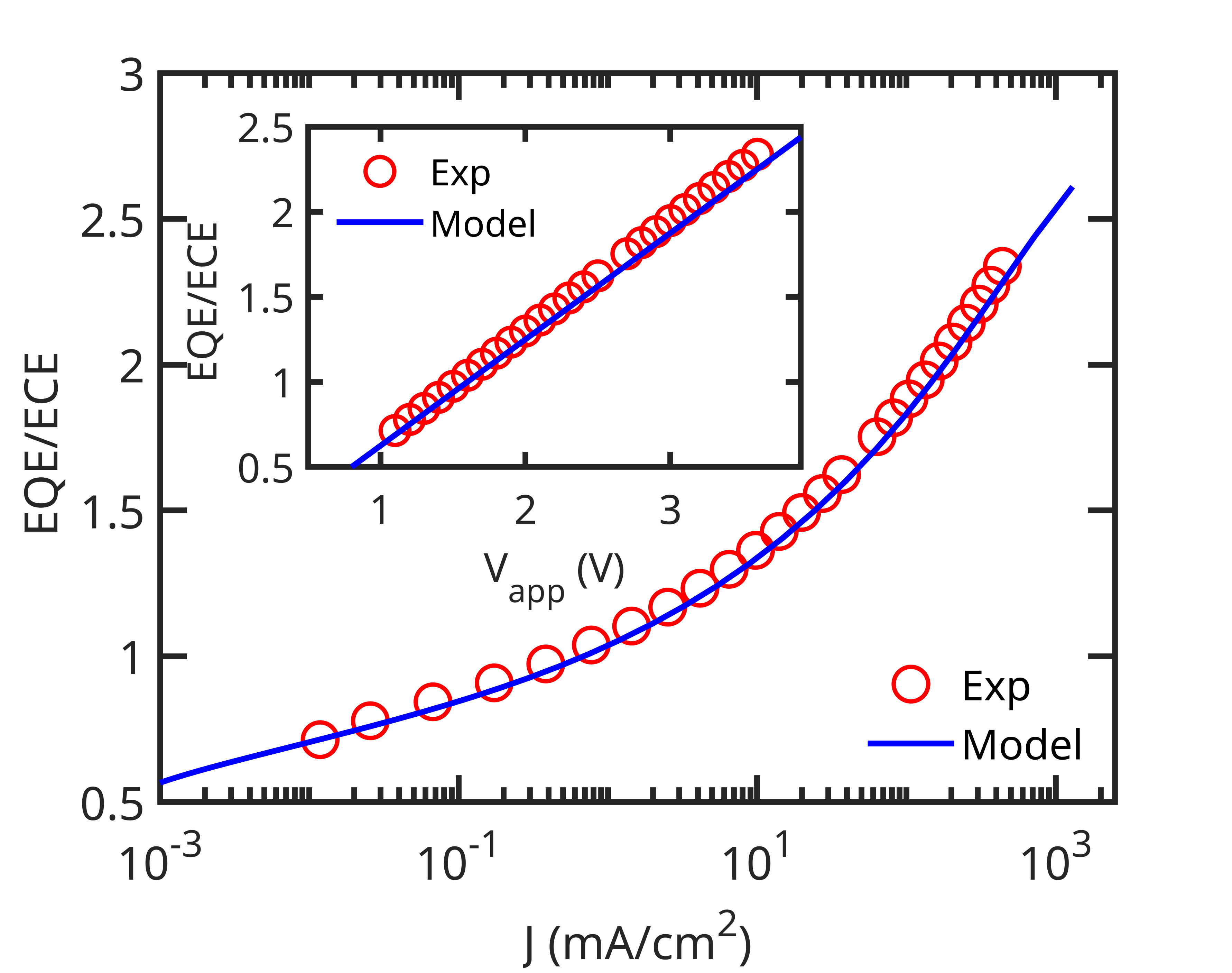}
    \caption{\textit{Variation of EQE/ECE as a function of $J$. The experimental results (symbols) are from Li et al.\cite{Li2024}. The solid lines are model predictions. The inset shows that the EQE/ECE varies linearly with $V_{app}$ with a slope proportional to $E_g$. The parameters used in model predictions are available in SI.
     }}
\label{ratio}
\end{figure}

The effectiveness of the above formalism to predict PeLED characteristics is illustrated in Figure \ref{ratio}. Here the ratio EQE/ECE is plotted as a function of $J$. The solid lines indicate model predictions while the symbols represent experimental results from a recent publication\cite{Li2024} on PeLEDs with maximum EQE of $32\%$ and maximum ECE of $25\%$. The inset of Figure \ref{ratio} shows the variation of EQE/ECE as a function of $V_{app}$. Eqs. \ref{eq:eqe} and \ref{eq:ece} lead to $\mathrm{EQE}/\mathrm{ECE} \propto V_{app}/E_g$ - i.e., $\mathrm{EQE}/\mathrm{ECE}$ depends only on $V_{app}$ which explains the trends in the inset of Figure \ref{ratio}. However, $J$ is a function of $V_{app}$ and is influenced by carrier recombination and space charge effects (see eqs. \ref{eq:Jrec}-\ref{eq:Vsc}). We find that the model well anticipates the experimental results in Figure \ref{ratio}.\\

The model system described by eqs. \ref{eq:Jrec}-\ref{eq:ece} well anticipates the experimental results on PeLEDs with a broad range of band gaps, active materials, and processing conditions\cite{nair2025multiphysics}. The same model with thermal transport and temperature dependent carrier recombination for efficiency roll-off under high injection conditions as well \cite{nair2024acs,nair2025multiphysics}. Comparison of model predictions with JV characteristics of multiple experimental data from the literature is provided in the  Supporting Information SI (see Figure S1). The calibrated model allows us to explore the sub $E_g$ operation limits of PeLEDs, as detailed below. The parameters involved in the model can be back extracted through measurements such as transient optical spectroscopy\cite{nair_exciton,nair2023characterization}, open circuit voltage transients\cite{Hossain2024}, dark JV analysis\cite{nair_jpcl2014,nair2025multiphysics}, etc.\\

\section{Limits of Ultra low voltage operation}

The low voltage operation prospects are often compared in terms of the turn-on voltage ($V_T$). There exist different definitions for the same. For example, Zheng et al., used working definition of $V_T$ as the applied voltage at which the luminance is $1\,\mathrm{cd/m^2}$ (see ref. \cite{zheng2024ultralow}). On the other hand, Lian et al. \cite{lian2022ultralow} defined $V_T$ as the applied voltage at which the emitted photon count is $10^9\,\mathrm{m^{-2}s^{-1}}$ - one of the most sensitive measurements reported so far. The same reference also reported light emission at sub $E_g$ voltages from several types of LEDs. However, in the absence of theoretical estimates, it remains unclear whether such findings truly represent the limits. Here we bridge this conceptual gap by elucidating the fundamental limits for sub $E_g$ operation for PeLEDs.\\

For sub-band gap operation, the current levels are typically low and hence $V_{app} \approx V_D$ (i.e., space charge effects are insignificant at low current levels). For a limiting threshold of detection of emitted photon flux $N_{T}$ (say, $10^9\,\mathrm{m^{-2}s^{-1}}$ as suggested by reference Lian et al.\cite{lian2022ultralow}), eqs. \ref{eq:VD} and \ref{eq:Nph}  indicate that the minimum applied voltage is 
\begin{equation}
     V_{T} = \frac{k_BT}{q}ln \left( \frac{N_T}{k_{2}n_i^2W_P\times \eta_{OC} } \right) \\
    \label{eq:VT1}
\end{equation}
In terms of $E_g$ and the density of states of the conduction and valence band (i.e., $N_C$ and $N_V$, respectively), we have
\begin{equation}
     V_{T} = E_g+\frac{k_BT}{q}ln \left( \frac{N_T}{k_{2}N_CN_VW_P\times \eta_{OC} } \right) \\
    \label{eq:VT2}
\end{equation}
The analytical estimates for $V_T$ yield several quantitative insights about sub $E_g$ operation of PeLEDS. For example, we find that $N_T< k_{2}N_CN_VW_P\times \eta_{OC}$ for typical parameters such as  $N_C=N_V=10^{19}\,\mathrm{cm^{-3}}$, $\eta_{OC} =0.36$, $W_P=60\,\mathrm{nm}$, $k_2=5\times 10^{-11}\,\mathrm{cm^3/s}$, and $N_T=10^9\,\mathrm{m^{-2}s^{-1}}$. This indicates that sub band gap operation is indeed possible with PeLEDs. Further, as the parameters in the second term in the RHS of eq. \ref{eq:VT2} are not strongly dependent on $E_g$, we expect $E_g-V_T$ to be nearly constant over a broad range of $E_g$. The reduced sensitivity of material parameters can be illustrated as follows. Eq. \ref{eq:VT2} indicates that, for a given $N_T$, even a two order of magnitude change in $k_2N_CN_VW_P\eta_{OC}$ leads to $\sim 120\,\mathrm{mV}$ change in $V_T$. Hence, a near constant offset between $V_T$ and $E_g$ is expected for multiple LED technologies. In addition, the sub $E_g$ operation limits, when expressed as a fraction of $E_g$, are expected to be lower for PeLEDs with lower $E_g$. \\

\begin {figure} [h!]
  \centering
    \includegraphics[width=0.45\textwidth]{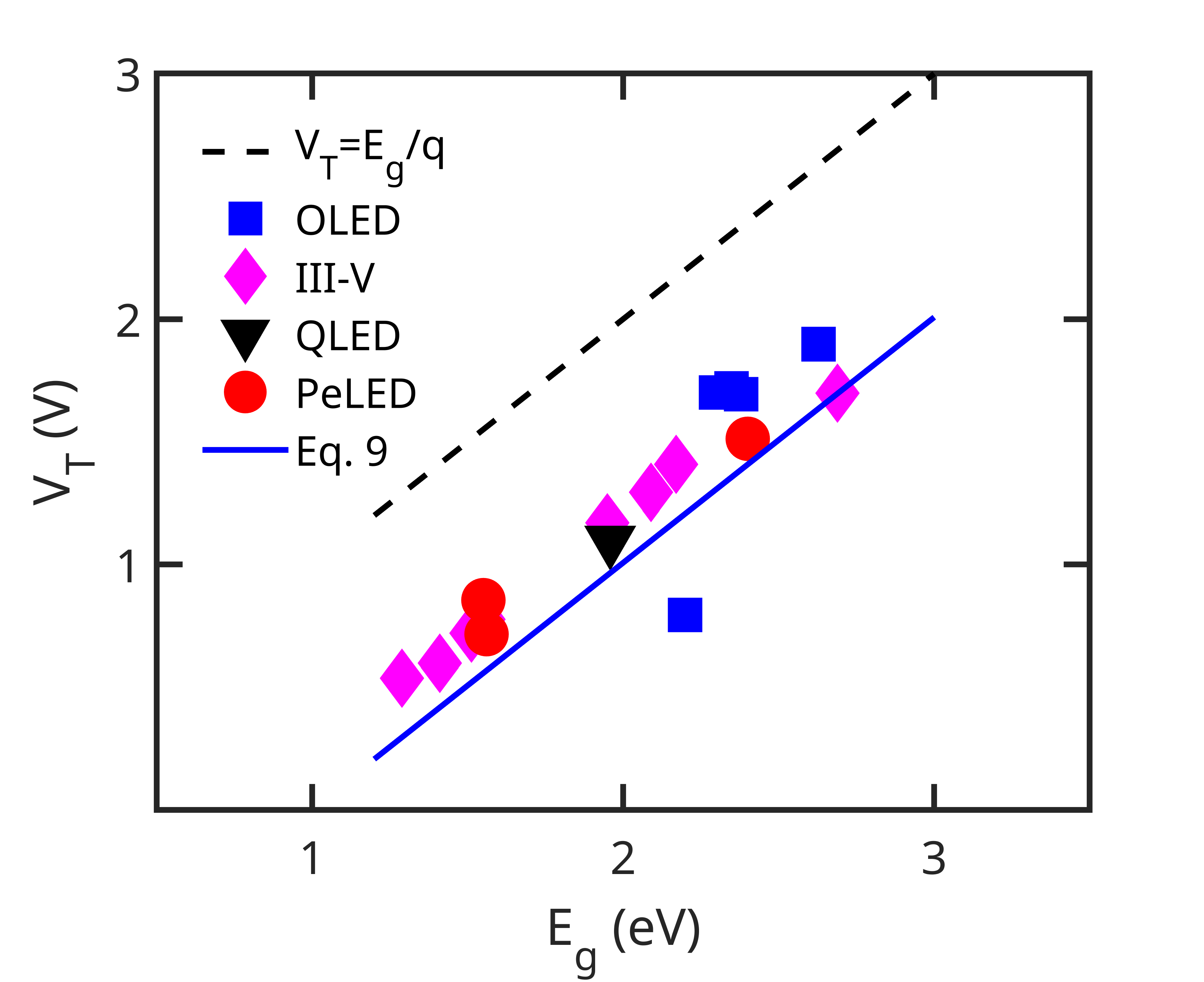}
    \caption{\textit{ Comparison of model predictions with experimental results on ultra low voltage operation of LEDs. The experimental results are from Lian et al.\cite{lian2022ultralow} for various types of LEDs as denoted in the legend. The theoretical prediction (solid line) correspond to parameters as mentioned in text. As anticipated by the model, $E_g-V_T$ is nearly independent of $E_g$ for a broad class of LED systems. The parameters used in model predictions are available in SI.
     }}
\label{VT_fig}
\end{figure}

Figure \ref{VT_fig} compares the above predictions with experimental results reported by Lian  et al.\cite{lian2022ultralow} In accordance with the theoretical predictions, we find that the experimental $V_T$ scales linearly with $E_g$. Further, $E_g-V_T$ is nearly the same for a broad range of $E_g$ which is in complete agreement with model predictions. This is a remarkable result as the experimental results plotted in Figure \ref{VT_fig} belong to different categories like organic LEDs (OLEDs), PeLEDs, quantum dot LEDs (QLED), and III-V based LEDs with a wide variation in their material and recombination parameters. Still, all of them show a linear variation with $E_g$ and a nearly constant offset - which is independent of $E_g$. In this regard, we note that eq. \ref{eq:VT2} is a general result and is independent of the assumption $n=p$ used in the model description of JV characteristics of PeLEDs. This is due to the fact that radiative recombination is proportional to $np$ which scales as $e^{qV_{app}/k_BT}$ for low bias - regardless of the assumption $n=p$. Hence, eq. \ref{eq:VT2} is expected to hold for a broad class of LEDs, as evident in Figure \ref{VT_fig}.\\

Alternate definition for ultra low voltage operation of PeLEDs is also available in literature. For example, Zheng et al. defined $V_T$ as the voltage at which the emitted radiance is $1\,\mathrm{cdm^{-2}}$. The  power density of emitted photons from a PeLED is given as $P_{out} = qE_gN_{ph}$. The radiance of a LED is directly proportional to $P_{out}$ with the scaling factor being the effective solid angle of emission \cite{huang2020mini}. Using the same set of parameters, we find that the predicted $V_T$ is about $1.8\,\mathrm{V}$ for $E_g \approx 2.35 \,\mathrm{eV}$ - in close agreement with the experimental results reported by Zheng et al.\cite{zheng2024ultralow} and Wang et al.\cite{wang2025efficient}

\section{Maximum EQE at sub $E_g$ bias? }
Theoretical estimates presented in the previous section resolved several puzzles regarding the limits for sub bandgap operation of PeLEDs. However, as defined, $V_T$ is not a fundamental quantity that can be expressed solely in terms of the device parameters. Indeed, the definition of $V_T$ is from an operational perspective and is heavily influenced by instrumentation/detection capability. In addition, very low values of $V_T$ are of little practical significance as the corresponding light emission is also nominal. A more relevant limit would be the voltage at which a PeLED could achieve its maximum EQE. Is it possible to achieve maximum EQE at sub $E_g$ voltages? \\


Interestingly, the internal voltage at which PeLEDs exhibit maximum EQE can be succinctly defined as follows: With the definition of internal quantum efficiency $\mathrm{IQE}=k_2n^2/(k_1+k_2n^2+k_3n^3)$, we find that IQE maximizes at a carrier density $n_{qe}=\sqrt{k_1/k_3}$. As per eq. \ref{eq:Jrec}-\ref{eq:eqe}, the key parameters of a PeLED at its maximum EQE are given as
\begin{align}
     V_{D,QE} & =  E_g+\frac{kT}{q} ln(\frac{k_1}{k_3})-\frac{kT}{q} ln(N_CN_V)
     \label{eq:vdqe}\\
     J_{QE}&=q\frac{k_1}{k_3}\times (k_2+2\sqrt{k_1k_3})W_P
     \label{eq:jqe}
\end{align}
Eq. \ref{eq:vdqe} predicts that $V_{D,QE}$ scales linearly with $E_g$. Further, as the material parameters in the RHS of eq. \ref{eq:vdqe} are not strong functions of $E_g$, we expect the offset $E_g-V_{D,QE}$ to be independent of $E_{g}$. Remarkably, unlike $V_T$ in eq. \ref{eq:VT2}, $V_{D,QE}$ and $J_{QE}$, as  defined by eqs. \ref{eq:vdqe}-\ref{eq:jqe}, are fundamental quantities since they are uniquely defined only by the material/device parameters (and not bias conditions or instrumentation capabilities). \\

With good material quality, eq. \ref{eq:vdqe} indicates that it is indeed possible to achieve high emission even with sub band-gap voltages. For typical parameters like $k_1=10^5\,\mathrm{s^{-1}}$, $k_3=10^{-28}\,\mathrm{cm^6/s}$, and $n_i=0.5\times10^6\,\mathrm{cm^{-3}}$ (for $E_g = 1.6\,\mathrm{eV}$), we find $n_{qe} \approx 3\times 10^{16}\,\mathrm{cm^{-3}}$, and $V_{D,QE}\approx 1.27V$. However, the applied voltage at which maximum EQE happens is not $V_{D,QE}$. An additional potential drop is needed to support the space charge limited transport. Hence, as per eq. \ref{eq:Vsc}, the corresponding applied voltage is 
\begin{equation}
     V_{EQE} =  V_{D,QE}+K_{SC}J_{QE}^\alpha 
    \label{eq:Vappqe}
\end{equation}

\begin {figure} [h]
  \centering
    \includegraphics[width=0.45\textwidth]{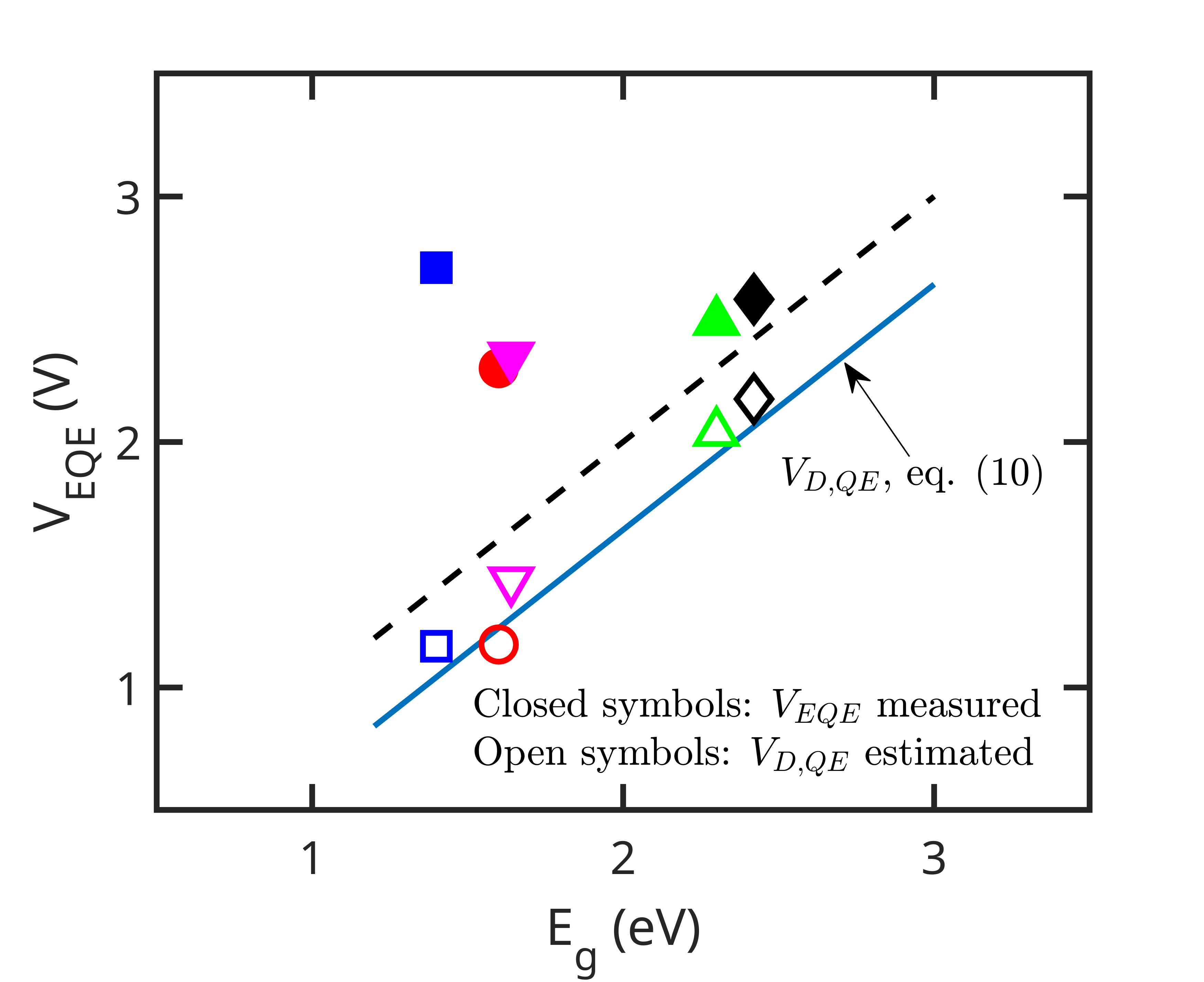}
    \caption{\textit{ Model predictions for PeLEDs to achieve maximum EQE at sub $E_g$ voltages. Here, the solid line denotes theoretical predictions as per eq. \ref{eq:vdqe}. The dashed line represents $V_{EQE}=E_g/q$. The solid symbols are experimental data from recent publications on high performance PeLEDs (red circle from Li et al.\cite{Li2024}, blue square from Jia et al.\cite{jia2021excess}, magenta triangle from Zhao et al.\cite{zhao2020thermal}, black diamond from Zheng et al.\cite{zheng2024ultralow}, and green triangle from Wang et al.\cite{wang2025efficient}). The open symbols denote estimated $V_{D,QE}$ as obtained through eqs. \ref{eq:vdqe}-\ref{eq:Vappqe}.  The parameters used in model predictions are available in SI.
     }}
\label{Vq_fig}
\end{figure}

Figure \ref{Vq_fig} compares the theoretical limits (solid line) with recent experimental data (solid symbols) from literature on high performance PeLEDs. The solid line indicates the absolute lower limits of applied voltage at which maximum EQE can be achieved (as per eq. \ref{eq:vdqe}). The experimental results (solid symbols) indicate that maximum EQE is achieved at applied biases much larger than $E_g$ for these high performance devices. The open symbols indicate the estimated $V_{D,QE}$ for these devices after accounting for the space charge effects (i.e., using eq. \ref{eq:Vappqe}, see SI for methodology). The theoretical limits are in good agreement with the $V_{D,QE}$ estimates from the experimental results. This indicates that once the transport limitations due to space charge effects are addressed, it is possible to achieve maximum EQE at sub $E_g$ applied biases. In addition, we also find that the experimental $E_g-V_{D,QE}$ is independent of $E_g$ in agreement with model predictions.\\ 

We note that the experimental $V_{EQE}$ from literature (solid symbols in Fig. \ref{Vq_fig}) show no apparent trends or correlation with $E_g$ (see item 'd' listed among the open questions in Section I). In fact, the experimental trends might even lead to an erroneous conclusion that $V_{EQE}$ is independent of $E_g$. However, our theoretical analysis allows consistent estimation of $V_{D,QE}$ from these experimental data (open symbols). The back extracted $V_{D,QE}$ shows clear scaling trends with $E_g$ as predicted by eq. \ref{eq:vdqe}. The methodology to back extract $V_{D,QE}$ from experimental $V_{EQE}$ is detailed in SI. This allows us to quantify the influence of space charge effects that lead to additional voltage drop and hence higher power consumption in PeLEDs. 

\section{Maximum ECE at sub $E_g$ bias? }
One of the important optimization challenges for LEDs is to maximize light output while minimizing power consumption. Hence, it is of natural interest to achieve maximum ECE conditions at low applied biases. In tune with EQE, is it possible to achieve maximum ECE at sub $E_g$ applied biases? It is interesting to note that maxima in EQE and ECE do not occur at the same bias conditions. This can be understood as follows: Using eq. \ref{eq:eqe}, we find that maximum EQE occurs under the condition 
\begin{equation}
 2n^{-1}\frac{\partial n}{\partial V_{app}}=J^{-1} \frac{\partial J}{\partial V_{app}}
 \label{eq:maxeqe}
\end{equation}
On the other hand, eq. \ref{eq:ece} indicates that maximum ECE occurs under the condition
\begin{equation}
 2n^{-1}\frac{\partial n}{\partial V_{app}}=J^{-1} \frac{\partial J}{\partial V_{app}}+\frac{1}{V_{app}}
 \label{eq:maxece}
\end{equation}
Equations \ref{eq:Jrec}-\ref{eq:Vsc} indicate that both $n$ and $J$ increase monotonically with $V_{app}$. Further, it can be shown that for low biases, $J^{-1}\partial J/\partial V_{app} \le 2n^{-1}\partial n/\partial V_{app}$. Together with eqs. \ref{eq:maxeqe}-\ref{eq:maxece}, these relations lead to  
\begin{equation}
V_{ECE}<V_{EQE}
\label{eq:vrel}
\end{equation}
which indicates that the maximum ECE occurs at a lower voltage than the maximum of EQE. Eqs. \ref{eq:vdqe}-\ref{eq:Vappqe} indicate that maximum EQE can be achieved at sub $E_g$ conditions (subject to improvements in space charge transport). Hence, it is evidently possible to achieve maximum ECE at sub $E_g$ applied biases.\\

\begin {figure} [h]
  \centering
    \includegraphics[width=0.45\textwidth]{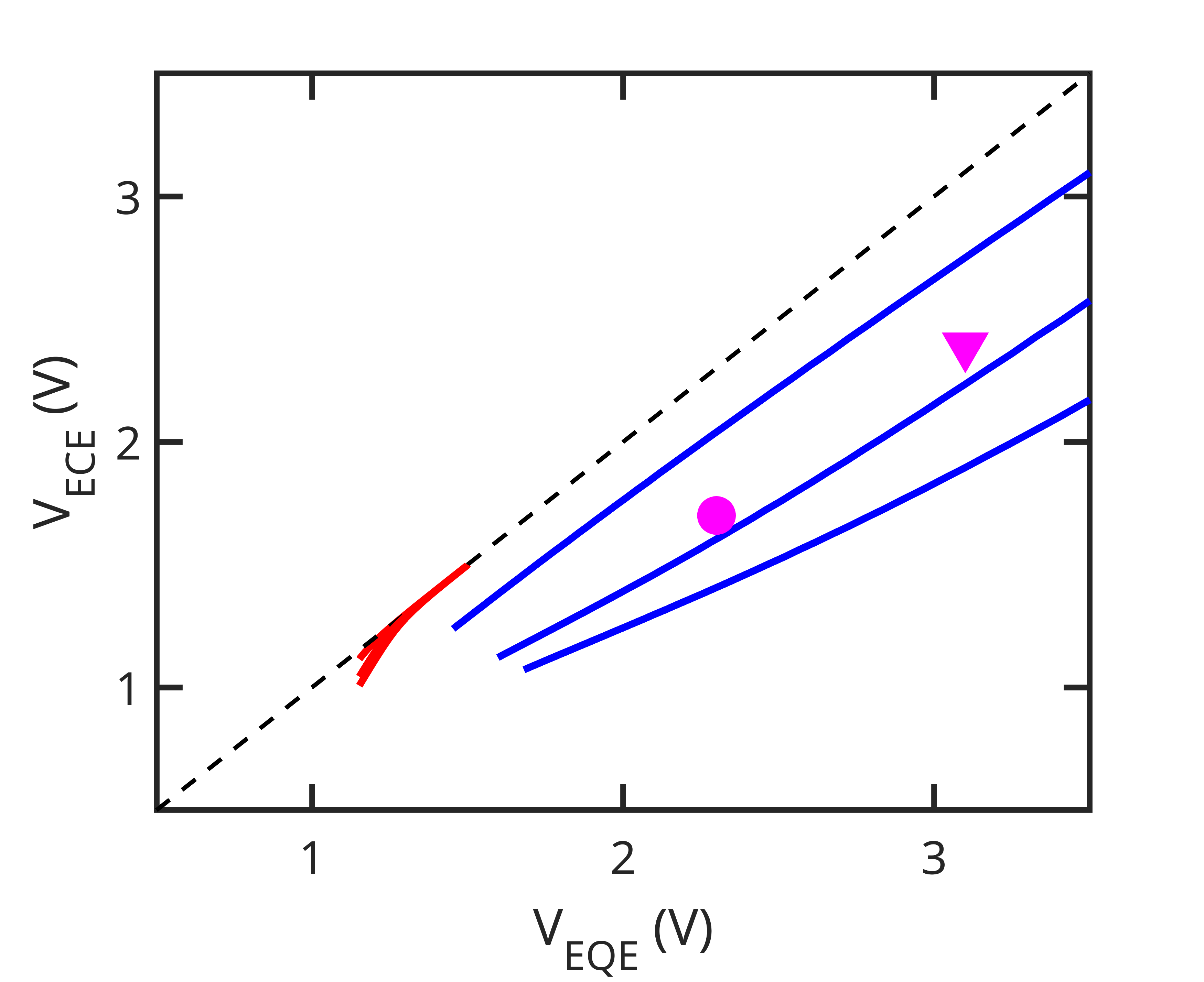}
    \caption{\textit{ Comparison between $V_{ECE}$ and $V_{EQE}$ for PeLEDs (for $E_g=1.6\,\mathrm{eV}$). The dashed line represent $V_{EQE}=V_{ECE}$ conditions while the solid lines denote model predictions as $k_1$ is varied for various listed $k_2$. The solid symbols are experimental data from recent literature (circle from Li et.\cite{Li2024} and triangle from Jia et al.\cite{jia2021excess}). Here, we find that $V_{ECE}< V_{EQE}$ as predicted by the theoretical model. The parameters used in model predictions are available in SI.
     }}
\label{Vqevce_fig}
\end{figure}

Figure \ref{Vqevce_fig} shows the trends for estimated $V_{EQE}$ and $V_{ECE}$ for various parameter combinations.The red lines indicate the trends for recombination limited operation of PeLEDs while the blue lines show the trends for space charge limited conditions. It is evident that for all conditions, $V_{ECE}\le V_{EQE}$ - in accordance with theoretical expectations. In addition, the available experimental data are in agreement with these predictions. Hence, it is important to reduce space charge effects to achieve sub $E_g$ values for $V_{ECE}$.\\

\section{Discussions}
The results in the previous sections identified the limits for low voltage operation for PeLEDs. Our model predicts the possibility of sub $E_g$ operation at maximum EQE and ECE conditions. In addition, the model identifies the limits for ultra low voltage operation of PeLEDs. As the photon detection capability improves, limits for such ultra low voltage operation could be achieved at even lower applied biases, as indicated by eqs. \ref{eq:VT1}-\ref{eq:VT2}. As such, there are no absolute lower limits for photon detection from LEDs. Interestingly, basic arguments on power conservation in LEDs lead to such an insight, as follows: 
The power consumption (per unit area) of a PeLED under forward bias conditions is $JV_{app}$, while the net power that radiates from the LED is given as $J\times \mathrm{EQE}\times E_g$. Hence, under steady state conditions, conservation of power leads to 
\begin{equation}
     J\times V_{app} >  J\times EQE\times E_g \\
    \label{eq:conserv}
\end{equation}
Consequently, we identify the following important theoretical constraints for both $V_{app}$ and $EQE$
\begin{subequations}
\begin{align}
     V_{app} &> \mathrm{EQE} \times E_g \\
    EQE  &< \frac{V_{app}}{E_g}
    \end{align}
    \label{eq:limits}
\end{subequations}
We note that EQE depends on the carrier density $n$ and therefore on both $J$ and $V_{app}$. Accordingly, eq. \ref{eq:limits}a is a lower limit for $V_{app}$ and indicates that it is possible to achieve photon emission (however negligible it may be) at any $V_{app}>0$ from a LED. Similarly, eq. \ref{eq:limits}b indicates that achievable EQE is always limited by $V_{app}$ and $E_g$. For example, an EQE of $30\%$ can be achieved only with applied biases larger than $0.3E_g$. In addition, eq. \ref{eq:limits}b indicates that the EQE can  increase as $V_{app}$ increases (i.e., from $V_{app}=0$) - as observed in almost all experiments. Further, the same equation also predicts that it is possible to achieve maximum EQE conditions at sub $E_g$ applied biases - which is yet to experimentally reported for PeLEDs. However, EQE roll-off under high injection conditions is not directly evident from the trends given in eq. \ref{eq:limits}, but the same was elaborated in detail in our recent publications\cite{nair2024acs,nair2025multiphysics}.\\

The results shared in this manuscript also identifies important optimization routes to achieve low power operation of PeLEDs. Eq. \ref{eq:vdqe} indicates that it is essential to lower $k_1$ to achieve low voltage, high radiance operation of PeLEDs. Further, it is beneficial to have materials with large $N_C, N_V$. However, we note that the estimated $V_{D,QE}$ might still be larger than the $V_{BI}$ of practical PeLEDs (which is of the order of 1V). As mentioned earlier, the dark J-V characteristics of PIN diodes vary exponentially with $V_{app}$ in the low bias regime till $V_D < V_{BI}$. In high bias regime, our results indicate that $V_{D}$ increases sub-linearly with $V_{app}$. Under such conditions, significant potential drop is required to support the space charge limited transport. This observation implies that the minimum operating voltage for practical devices could be more than $V_{D,QE}$. Such large $V_{app}$ leads to undesired power consumption and efficiency roll-off as well. Hence, to achieve low voltage high radiance operation of PeLEDs, we need to: (a) reduce mono-molecular recombination, (b) increase the conductivity of transport layers, and (c) increase the $V_{BI}$ of the device. \\

We used a semi-classical band description to model the PeLEDs in this manuscript. The same has been shown to be appropriate to describe a wide variety of perovskite based optoelectronic devices. The material parameters like recombination coefficients, density of states, etc. influence the limiting biases defined in eqs. \ref{eq:VT2} and \ref{eq:vdqe}. However, as the dependence is logarithmic in nature, we find that the same equations predict trends from LEDs of diverse technologies.  It is well known that mobile ions could impact the performance of such devices. The assumption $n=p$ is appropriate for devices with large concentration of mobile ions as they screen the internal electric field\cite{reenen_jpcl,saketh2021ion,sivadas2023ionic,sivadas2021efficiency}. Literature also indicates that a band level description can often be useful for Organic devices\cite{koster2005device,ray2011annealing,ray2015collection}. Consequently, OLEDs also follow the general rule given by eq. \ref{eq:VT2}. However, we notice that the experimental $V_T$ for one OLED device is an exception and is much lower than the rest of the devices (see Figure \ref{VT_fig}). This OLED is based on Rubrene, whose electroluminescence could be dominated by triplet-triplet annihilation \cite{chen2016determining}. Our model neglects trap filling effects which could be significant at low bias regime of the JV characteristics\cite{kim2022ultra}. However, this is not a drawback as significant light emission happens, typically, beyond the trap filling voltages. If need be, such trap filling effects could be incorporated in our model through additional terms, as appropriate. We also notice the possibility of a novel characterization scheme for LEDs - the band gap can be directly measured from the EQE/ECE vs. $V_{app}$ plot, provided EQE and ECE are obtained through independent measurements (see inset of Figure \ref{ratio}).  

\section{Conclusions}
To summarize, here we propose a theoretical framework to explore the possibilities of low voltage, high radiance operation of PeLEDs. Our model well anticipates experimental reports on ultra low voltage light emission from diverse devices like OLEDs, PeLEDs, QLEDs, and III-V based LEDs. In addition, we identify the prospects to achieve maximum EQE and ECE at sub $E_g$ voltages for PeLEDs. Analyses of experimental results from recent reports on high performance PeLEDs help us identify important optimization pathways to achieve high radiance low voltage operation. The insights shared and methodology adopted in this manuscript could be of broad interest to several classes of LEDs.

\section{Acknowledgement}
PRN acknowledges National Center for Photovoltaics Research and Education (NCPRE), Indian Institute of Technology Bombay and discussions with Adwaith Kiran Marathi, Simhadri Venkata Ramana, and Sushma Usurupatti, IIT Bombay.\\

\section*{References}
\bibliography{references}

\end{document}